\documentclass[doublecol]{epl2} 
\usepackage{graphicx}
\usepackage{amsmath,amssymb,dsfont}

\usepackage{xcolor}
\usepackage{wrapfig}
\usepackage{lipsum}

\usepackage{hyperref}
\hypersetup{
    colorlinks=true,
    linkcolor=blue,
    filecolor=blue,      
    urlcolor=blue,
    citecolor=blue
    }

\usepackage{subfig}
\begin{document}
\title{Discrete Schr\"odinger equation on graphs:\\ An effective model for branched quantum lattice}
\author{M. Akramov\inst{1},
C. Trunk\inst{2}, J. Yusupov\inst{3},
D. Matrasulov\inst{4,5}}

\shortauthor{M. Akramov \etal}
\institute{
\inst{1} National University of Uzbekistan, Universitet Str. 4, 100174, Tashkent, Uzbekistan, \url{mashrabresearcher@gmail.com}\\
\inst{2} Institute for Mathematics, TU Ilmenau, Weimarer Str. 25, 98693, Ilmenau, Germany, \url{carsten.trunk@tu-ilmenau.de}\\
\inst{3} Kimyo International University in Tashkent, Usman Nasyr Str. 156, 100121, Tashkent, Uzbekistan, \url{jambul.yusupov@gmail.com}\\
\inst{4} Turin Polytechnic University in Tashkent, Niyazov Str. 17, 100095, Tashkent, Uzbekistan, \url{dmatrasulov@gmail.com}\\
\inst{5} Center for Theoretical Physics, Khazar University, 41 Mehseti Street, Baku, AZ1096, Azerbaijan
}

\abstract{
We propose an approach to quantize discrete networks (graphs with discrete edges). We introduce a new exact solution of discrete Schr\"odinger equation that is used to write the solution for quantum graphs. Formulation of the problem and derivation of secular equation  for arbitrary quantum graphs is presented. Application of the approach  for the star graph is demonstrated by obtaining eigenfunctions and eigenvalues explicitely.  Practical application of the model in conducting polymers and branched molecular chains is discussed.}
\maketitle

\section{Introduction}
Modeling of quantum particle motion in low-dimensional branched structures is of importance for many practical applications, especially for the designing and optimization of functional properties of quantum materials. A powerful tool for such purpose is to use quantum graph theory. The advantage of quantum graphs in modeling quantum transport in low-dimensional branched structures and networks comes from the fact that the description can be effectively reduced to a one-dimensional Schr\"odinger (Dirac) equation on metric graphs. In most of the cases, quantum graphs based approach allows one to obtain an exact solution to the problem for arbitrary graph topologies. 

Quantum graphs are determined as the branched quantum wires, which are connected to each other at the nodes (vertices)  according to a certain rule called the topology of a graph. 
Initially, the quantum graphs were introduced for modeling of electron motion in low-dimensional molecular structures in the quantum chemistry of organic molecules. In particular, the Refs.~\cite{Pauling, Rud, Alex}, where the electron motion in branched aromatic molecules was studied, can be considered as a pioneering attempt. However, the strict formulation of the quantum graph concept, where the latter was defined as branched quantum wire, has been presented a few decades later by Exner and Seba in \cite{Exner1}. The next step in the systematic study was done by Kostrykin and Schrader, who proposed general vertex boundary conditions ensuring self-adjointness of the Schr\"odinger operator on graphs \cite{Kost}. Later the
quantum graph concept has been used in different contexts (see, the Refs.\cite{Uzy1, Exner15, Hul,  PTSQGR}) and an experimental realization in microwave networks was done \cite{Hul}.  For an overview of different mathematical aspects of the quantum graph theory, we refer to the books \cite{Exner15, Grisha, Kurasov}. Some physically important problems of the quantum graphs theory are studied in the Refs.\cite{Jambul,JM,KarimBdG,Alex1,PQG2,PQG6,Rabinovich,Grisha2,PQG8}. Different issues of evolution and spectral equations on fractal and discrete systems, including graphs are considered in \cite{Rezapour1,Rezapour2,Rezapour3,Rezapour4,Rezapour5,Rezapour8,Rezapour9,Rezapour10,Rezapour11}.

In this paper, we consider a version of a quantum graph with discrete edges, i.e., each branch represents a discrete chain of finite length. Such a system models branched lattices, where particle motion occurs in the quantum regime. 

Using the solution of the discrete Schr\"odinger equation on a finite 1D lattice, we construct a solution of the problem for the discrete quantum graph, which satisfies vertex boundary conditions. Finally, the eigenvalues are determined in terms of a secular equation.

\section{Exact solution and spectrum of discrete Schr\"odinger equation on a finite chain}\label{dse_on_aline}

Discrete Schr\"odinger equation has attracted attention in the context of mathematical physics, exactly solvable models and spectral theory \cite{DSE2,DSE1,PQG7,PQG5}.
The one-dimensional time-independent Schr\"odinger equation for a free particle is written as
\begin{equation}\label{eq1}
\frac{d^2\Psi(x)}{d x^2}+k^2\Psi(x)=0,
\end{equation}
on the infinite line $x\in(-\infty,+\infty)$, where $\hbar = 2m = 1$ and $\Psi(x)$ is the wave function. The discrete version of the Schr\"odinger equation can be obtained by using the standard finite-difference scheme with the step size $a>0$ for the second-order derivative as
\begin{align}\label{eq2}
\begin{split}
\frac{1}{a^2}\bigg[\Psi^{(a)}(x_n^{(a)}-a)-2\Psi^{(a)}(x_n^{(a)}&)+\Psi^{(a)}(x_n^{(a)}+a)\bigg]\\
&+k^2\Psi^{(a)}(x_n^{(a)})=0,
\end{split}
\end{align}
where $x_n^{(a)}=na$ and $n$ is an integer number.

Here we propose an exact solution of Eq.~\eqref{eq2} which can be written as
\begin{equation}\label{sol1}
\Psi^{(a)}(x_n^{(a)})=A g_+(a)^{x_n^{(a)}/a}+B g_-(a)^{x_n^{(a)}/a},
\end{equation} 
where $A$ and $B$ are constants and
\begin{equation*}
g_{\pm}(a)=1+\frac{-k^2 a^2 \pm ka\sqrt{k^2 a^2-4}}{2}.    
\end{equation*}

\begin{figure}[t!]
    \centering
    \includegraphics[width=85mm]{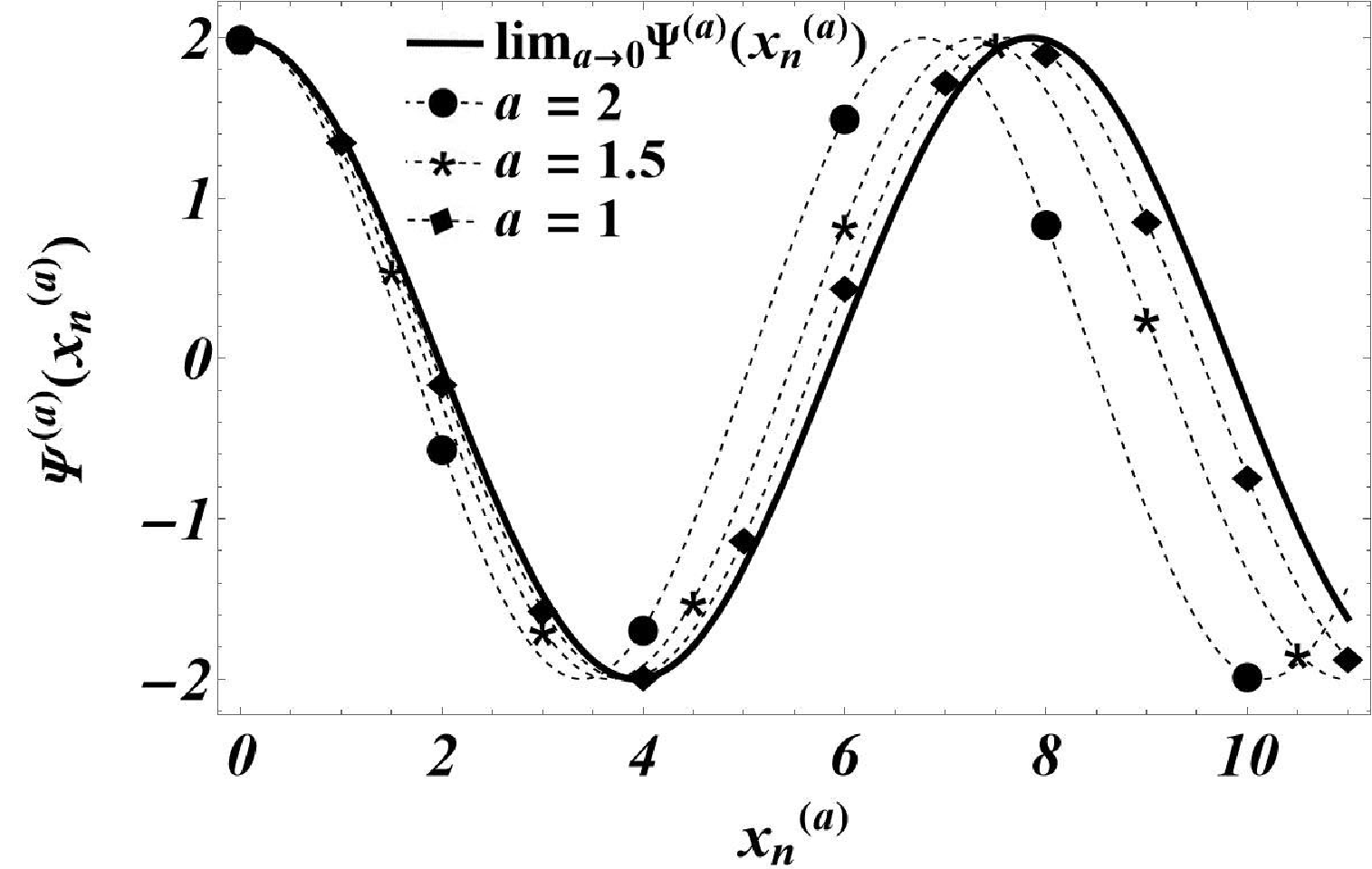}
    \caption{Analytic solution of discrete Schr\"odinger equation in Eq.~\eqref{sol1} for different values of the step size $a$ and limit of the solution in Eq.~\eqref{limit} for parameters $k=0.8$, $A=B=1$.}
    \label{fig:solution}
\end{figure}

The consistency of this solution with the solution of the continuous case can be shown by considering its limit for $a\to 0$ for a fixed point $x_n^{(a)}$. For brevity
we set
\begin{equation*}
    z_{\pm}:=\frac{-k^2 a^2 \pm  ka \sqrt{k^2 a^2-4}}{2}
    \quad \mbox{and}\quad \alpha :=  x_n^{(a)}.
\end{equation*}
As $z_\pm$ tends to zero for $a\to 0$, we obtain 
\begin{align}
\begin{split}
    \lim_{a\to 0}\left( g_{\pm}(a) \right)^{\alpha/a} &= 
    \lim_{a\to 0}\left( 1+z_{\pm} \right)^{\alpha/a} \\
    &=
    \lim_{a\to 0}\left[\left( 1+z_{\pm} \right)^{1/z_{\pm}}\right]^{\alpha z_{\pm}/a}\\ 
    &=
    \exp{\left(\lim_{a\to 0} \alpha z_{\pm}/a\right)} = 
    e^{\pm ik\alpha}.
    \end{split}
\end{align}
One can easily show that
\begin{align} \label{limit}
\begin{split}
\lim_{a\to 0} \Psi^{(a)}(\alpha) = 
\lim_{a\to 0} 
\bigg(
A[g_+(a)]&^{\alpha/a}+
B[g_-(a)]^{\alpha/a}
\bigg) \\
&= Ae^{ik\alpha}+Be^{-ik\alpha}.
\end{split}
\end{align}
Graphical representation of the convergence of the solution \eqref{sol1} to the solution of the continuous case with a decrease of the step size $a$ is shown in Fig.~\ref{fig:solution}.

\begin{table*}[t] \caption{The first five nonzero eigenvalues $k$ for Dirichlet (left) and Neumann (right) boundary condition compared with the continuous case.}
    \parbox{.49\linewidth}{
        \centering
        \begin{tabular}{l*{3}{|c}r}
Continuous     &  $a=0.1$   & $a=0.01$  & $a=0.002$ \\
case          &  $N=10$    &  $N=100$  & $N=500$ \\
\hline
 $\pi$     & 3.1286893  & 3.1414634 & 3.1415874\\
 $2\pi$    & 6.1803398  & 6.2821518 & 6.2831439\\
 $3\pi$    & 9.0798099  & 9.4212901 & 9.4246384\\
 $4\pi$    & 11.7557050 & 12.5581039& 12.5660398\\
 $5\pi$    & 14.1421356 & 15.6918191& 15.7073173
\end{tabular}}
    \hfill
    \parbox{.49\linewidth}{
        \centering
        \begin{tabular}{l*{3}{|c}r}
Continuous     & $a=0.1$    & $a=0.01$   &  $a=0.002$ \\
case         &  $N=10$    &  $N=100$   &  $N=500$  \\
\hline
 $\pi$     & 3.4729635  & 3.1731927  &  3.1478832 \\
 $2\pi$    & 6.8404028  & 6.3455866  &  6.2957352 \\
 $3\pi$    & 9.9999999  & 9.5163831  &  9.4435249 \\
 $4\pi$    & 12.8557521 & 12.6847839 &  12.5912209\\
 $5\pi$    & 15.3208888 & 15.8499913 &  15.7387923 
\end{tabular}}
\label{eigenvalues}
\end{table*}

Now let's consider the problem on the finite discrete interval $[0, L]$ with step size $a=L/N$ for some 
$N \in \mathbb{N}$ and $x_n^{(a)}=na$ for $n=0,..., N$. 
Imposing Dirichlet boundary condition as 
\begin{eqnarray}\label{dirichlet}
    \Psi^{(a)}(0)=0, \quad \Psi^{(a)}(Na)=0
\end{eqnarray} 
leads to the following homogeneous system of linear equations for $A$ and $B$:
\begin{align}
\begin{split}
&A+B=0,\\ 
&A g_+(a)^{N}+B g_-(a)^{N} = 0.
\end{split}
\end{align}
The existence of non-trivial solutions of the above algebraic system leads to the following secular equation with respect to $k$
\begin{equation}\label{sec1}
        g_+(a)^N = g_-(a)^N.
\end{equation}

First, we consider the case $k^2a^2>4$ and $k>0$. Then 
\begin{equation*}
    g_+(a)-g_-(a)=2ka\sqrt{k^2a^2-4}>0 
\end{equation*}
and $g_+(a)>g_-(a)$ follows. Hence \eqref{sec1} does not hold.
A similar arguments shows the same for the case $k^2a^2>4$ and $k<0$.

It remains to consider the case $k^2a^2\leq 4$. Then 
\begin{eqnarray}
    g_{\pm}(a)=1+\frac{-k^2 a^2 \pm ika\sqrt{4-k^2 a^2}}{2}.
\end{eqnarray}
We write $g_{\pm}(a)$ as
\begin{eqnarray}
    g_{\pm}(a)=|g_{\pm}(a)|e^{i\varphi_{\pm}(a)},
\end{eqnarray}
and we have
\begin{align*}
    &\text{Re}(g_{\pm}(a)) = 1-\frac{k^2a^2}{2}, \\ 
    &\text{Im}(g_{\pm}(a)) = \pm\frac{1}{2} ka \sqrt{4-k^2a^2}.
    \end{align*}
    As $|g_{\pm}(a)|=1$, we obtain
\begin{equation}\label{Cabanyal}
\begin{split}
    & \varphi_{+}(a) = \left\{\begin{matrix}
        \arccos\left( 1-k^2a^2/2 \right), & \text{if } k>0,\\[0.5ex]
        2\pi-\arccos\left( 1-k^2a^2/2 \right), & \text{if } k<0,
    \end{matrix}\right. \\[1ex]
    & \varphi_{-}(a) = \left\{\begin{matrix}
        2\pi-\arccos\left( 1-k^2a^2/2 \right), & \text{if } k>0,\\[0.5ex]
        \arccos\left( 1-k^2a^2/2 \right), & \text{if } k<0.
    \end{matrix}\right.
    \end{split}
\end{equation}
By substitution of these equations into the secular equation \eqref{sec1} we 
see that $N\varphi_+(a)-N\varphi_-(a)$ equals a multiple of $2\pi$,
which gives by \eqref{Cabanyal}
\begin{equation*}
    \arccos\left( 1-k^2a^2/2 \right) = \frac{\pi m}{N},
\end{equation*}
where $m$ is an integer. Hence for every integer $m$ we obtain a solution
$k_m$,
\begin{eqnarray}
    k_m=\pm \frac{1}{a}\sqrt{2-2\cos\left(\frac{\pi m}{N}\right)}.\label{eigenvalues_discrete}
\end{eqnarray}
That is, equation \eqref{eq2} equipped with Dirichlet boundary 
conditions has solutions only if $k\in \{k_m \, :\, m\in \mathbb Z\}$.
Obviously, the $k_m$ depend also on the step size $a$.
If the step size $a$ tends to zero, then  $k_m$ converges to 
the eigenvalues of the continuous case on the interval $[0, L]$,
as the following calculation shows
\begin{align}\label{limit_eigenvalue}
\begin{split}
    \lim_{a\to 0} k_m = \pm&\sqrt{\lim_{a\to 0} \frac{2-2\cos\left(\frac{\pi ma}{L}\right)}{a^2} }\\
    &=\pm \frac{\pi m}{L} \sqrt{\lim_{a\to 0} \frac{\sin\left(\frac{\pi ma}{L}\right)}{\frac{\pi ma}{L}} }=\pm\frac{\pi m}{L}.
\end{split}
\end{align}

Analogously one can impose Neumann boundary condition as
\begin{equation}\label{Neumann}
    \Psi^{(a)}(a)-\Psi^{(a)}(0)=0, \quad \Psi^{(a)}(Na)-\Psi^{(a)}(Na-a)=0,
\end{equation}
which leads to the following homogeneous system of linear equations
\begin{align}
    A (g_+(a)-1 &) + B(g_-(a)-1)=0, \nonumber\\
    A(g_+(a)^N & - g_+(a)^{N-1})\\
    &+B(g_-(a)^N - g_-(a)^{N-1})=0. \nonumber
\end{align}
The secular equation in this case is written as
\begin{equation}\label{sec2}
    \det\begin{pmatrix}
        g_+(a)-1 & g_-(a)-1 \\
        g_+(a)^N - g_+(a)^{N-1} & g_-(a)^N - g_-(a)^{N-1}
    \end{pmatrix}=0.
\end{equation}

Substitution of the above definition of $g_{\pm}(a)$ to the determinant equation \eqref{sec2} yields to the following
\begin{eqnarray*}
    (e^{i\varphi_+(a)}-1)(e^{i\varphi_-(a)}-1) (e^{i\varphi_+(a)(N-1)}-e^{i\varphi_-(a)(N-1)})=0,
\end{eqnarray*}
or
\begin{eqnarray*}
    \varphi_{\pm}(a)=2\pi p, \quad 
    \varphi_+(a)-\varphi_-(a)=\frac{2\pi m}{N-1},
\end{eqnarray*}
where $p$ and $m$ are integer numbers.
Using the expression of $\varphi_{\pm}(a)$, one can derive the eigenvalues as
\begin{eqnarray}
    k_m=\pm \frac{1}{a}\sqrt{2-2\cos\left(\frac{\pi m}{N-1}\right)}. \label{eigenvalues_discrete2}
\end{eqnarray}
Similarly, it is easy to show that the convergence of the eigenvalues to the continuous case as shown for the Dirichlet boundary condition as
\begin{equation}\label{limit_eigenvalue2}
    \lim_{a\to 0} k_m = \pm\frac{\pi m}{L}.
\end{equation}

We collect the first five nonzero eigenvalues of continuous Schr\"odinger equation \eqref{eq1} with Dirichlet and Neumann boundary conditions defined on the interval $[0,1]$ which are given by $k_m=\pi m$, $m=1,...,5$ and compare them in Table \ref{eigenvalues} with the first five nonzero eigenvalues of the discrete Schr\"odinger equation \eqref{eq2} with Dirichlet \eqref{dirichlet} and Neumann \eqref{Neumann} boundary conditions on the interval $[0,1]$, respectively, for different values of step size $a$.

\section{Basic theory of quantum graphs}\label{cont_graph}

In this section, we will briefly recall the introduction of the quantum graphs and their spectrum by following the Ref.~\cite{Uzy1}.
A graph consists of $V$ vertices, where
$V$ is a natural number, and it is connected by $E$ edges, $E\in \mathbb N$. If vertices $i$ and $j$ are connected we call it
the $(i,j)$ edge.
Here we always assume that at most one edge
connects the same two vertices and moreover, that there are no loops, i.e.\ there is no
edge of the form $(i,i)$.

The topology of graph is defined by its $V\times V$ adjacency matrix as
\begin{eqnarray}
    C_{i,j}=C_{j,i}=\left\{
\begin{matrix}
    1, & \text{if} \; i \; \text{and} \; j \; \text{are} \; \text{connected,} \\
    0, & \text{otherwise,}
\end{matrix}
    \right.
\end{eqnarray}
where $i,j=1,2,3,..,V$. The number of edges can be found in terms of the adjacency matrix by $E=\frac{1}{2}\sum_{i,j=1}^V C_{i,j}$.

The coordinate $x_{i,j}$ on the edge $(i,j)$ is assigned as $[0,L_{i,j}]$ for all $C_{i,j}=1$ and $i<j$, where $L_{i,j}$ is the length of the edge $(i,j)$.
The wavefunction $\Psi$ is a vector function with $E$ components, where each component is defined as $\Psi_{i,j}(x_{i,j})$ for $C_{i,j}=1$.
The Schr\"odinger equation on each edge of the graph is written as (in units $\hbar = 2m = 1$)
\begin{equation}\label{schrodinger_graph}
\frac{d^2\Psi_{i,j}(x_{i,j})}{d x^2}+k^2\Psi_{i,j}(x_{i,j})=0.
\end{equation}
As usual, the Schr\"odinger operator $H=-\frac{d^2}{dx^2}$
on the graph is defined component wise on each edge of the graph
\begin{equation*}
\Psi_{i,j}(x_{i,j})\mapsto -\frac{d^2\Psi_{i,j}(x_{i,j})}{d x^2}.
\end{equation*}
Usually, one associates boundary conditions to 
the operator $H$. Without any boundary conditions, the Schr\"odinger operator $H$  satisfies 
\begin{align}\label{skew_Hermitian}
\begin{split}
    \langle H\Psi, \Phi\rangle - \langle \Psi,& H\Phi\rangle = \\
    \sum_{\substack{i,j=1 \\ i<j}}^V& C_{i,j}\left[  \Psi_{i,j}(x_{i,j}) \frac{d\Phi_{i,j}^*(x_{i,j})}{dx}\right.\\
    &\left. - \Phi_{i,j}^*(x_{i,j}) \frac{d\Psi_{i,j}(x_{i,j})}{dx}  \right]\bigg|_{x_{i,j}=0}^{x_{i,j}=L_{i,j}}.
\end{split}
\end{align}
Obviously, whenever the Schr\"odinger operator $H$ is self-adjoint, the expression in  \eqref{skew_Hermitian} is zero.
E.g., this is the case when the functions from the domain satisfy continuity 
on the graph. That is, for each vertex there exists a
complex number $\phi_i$, $i\in \{1, \ldots, V\}$, such that 
\begin{equation}\label{Valencia}
    \Psi_{i,j}(x_{i,j})\big|_{x_{i,j}=0}=\phi_i, \quad 
    \Psi_{i,j}(x_{i,j})\big|_{x_{i,j}=L_{i,j}}=\phi_j,
\end{equation}
for $i,j$ such that $C_{i,j}=1$, $i<j$, 
and some current conservation rule \cite{Uzy1},
\begin{align}\label{Spain}
\begin{split}
    -\sum_{j<i} C_{i,j} & \frac{d \Psi_{j,i}(x_{j,i})}{dx}\bigg|_{x_{j,i}=L_{j,i}} 
    \\
    &+\sum_{j>i} C_{i,j}
    \frac{d\Psi_{i,j}(x_{i,j})}{dx}\bigg|_{x_{i,j}=0} = \lambda_i \phi_i,
\end{split}
\end{align}
for some real parameters $\lambda_i$ for $i$ in $\{1,...,V\}$.
Note that the case $\lambda_i=0$ for all  $i \text{ in } \{1,...,V\}$ is called
Kirchhoff rule.

The solution of the Schr\"odinger equation on the graph in~\eqref{schrodinger_graph} can be written as
\begin{gather}\label{sol_continuous}
    \Psi_{i,j}(x_{i,j}) = C_{i,j}
    \left[  A_{i,j} \sin[k(L_{i,j}-x_{i,j})] + B_{i,j} \sin(kx_{i,j}) \right],
\end{gather}
for all $i<j$.
Substituting the solution \eqref{sol_continuous} into the vertex boundary conditions \eqref{Valencia} and \eqref{Spain} leads to the secular equations as
\begin{equation}\label{Tapas}
    \begin{split}
    &C_{i,j}A_{i,j}\sin(kL_{i,j}) = C_{i,j} \phi_i,\\ 
    &C_{i,j}B_{i,j}\sin(kL_{i,j}) = C_{i,j} \phi_j,
    \end{split}
\end{equation}
for $i,j$ such that $C_{i,j}=1$ and $i<j$, and
\begin{align}\label{Paella}
\begin{split}
    \sum_{j<i} C_{i,j} & k \left[ 
    A_{j,i} - B_{j,i} \cos(kL_{j,i})
    \right] \\
    +&\sum_{j>i} C_{i,j}k \left[ 
     -A_{i,j} \cos(kL_{i,j}) + B_{i,j}
    \right] = \lambda_i \phi_i,
\end{split}
\end{align}
for all $i$ in $\{1,...,V\}$.

This is a system of linear homogeneous equations for the unknowns 
$A_{i,j}, \; B_{i,j}$ and $\phi_i$ and has a 
non-trivial solution only when
\begin{eqnarray}\label{eigenvalues_continuous}
    \det(M(k))=0,
\end{eqnarray}
where $M(k)$ is a  matrix where all the coefficients of the linear 
equations \eqref{Tapas} and \eqref{Paella} are collected. It depends
on the eigenvalue parameter $k$ and its size is
\begin{equation*}
    (2E+V)\times (2E+V).
\end{equation*}

\section{Branched lattices}
\subsection{Arbitrary branching topology}
Adhering to the formulation of the quantum graph problem proposed in \cite{Uzy1}, in this subsection, we present a prescription for solving the problem for discrete quantum graphs.

The discrete graph is a set of $V$ vertices connected by $E$ discrete edges. We denote the edge which connects the vertices  $i$ and $j$ as $(i,j)$ for $i<j$.

For $(i,j)$ such that $C_{i,j}=1$ (i.e., vertex $i$ and vertex $j$ are connected) we assign discrete coordinates of the edge $x_0^{(a_{i,j})},...,x_{N_{i,j}}^{(a_{i,j})}$, which has step size $a_{i,j}$. Here $x_0^{(a_{i,j})}=0$  and $x_{N_{i,j}}^{a_{i,j}}$ is the right endpoint. In other words, each edge of the graph is an interval with discrete points starting at zero and ending at $x_{N_{i,j}}^{(a_{i,j})} = L_{i,j} = N_{i,j} a_{i,j}$.

The wavefunction $\Psi$ is a vector with $E$ components and each of its components is a vector in $\mathbb{C}^{N_{i,j}+1}$ defined on each edge $(i,j)$ as 
\begin{equation*}
    \left(\Psi_{i,j}^{(a_{i,j})}(x_{n_{i,j}}^{(a_{i,j})})  \right)_{n_{i,j}=0}^{N_{i,j}},
\end{equation*}
with $C_{i,j}=1$, where $i<j$.

The discrete Schr\"odinger equation on each edge is given by
\begin{align}
\begin{split}
    &\frac{1}{a_{i,j}^2}  [\Psi_{i,j}^{(a_{i,j})}(x_{n_{i,j}}^{(a_{i,j})}-a_{i,j})-2\Psi_{i,j}^{(a_{i,j})}(x_{n_{i,j}}^{(a_{i,j})}) \\
    &+\Psi_{i,j}^{(a_{i,j})}(x_{n_{i,j}}^{(a_{i,j})}+a_{i,j})]+k^2\Psi_{i,j}^{(a_{i,j})}(x_{n_{i,j}}^{(a_{i,j})})=0, \label{eq_graph}
\end{split}
\end{align}
where $1\leq n_{i.j} \leq N_{i,j}-1$ and $C_{i,j}=1$, $i<j$.

We use the standard Euclidean inner product on the discrete graph
\begin{eqnarray}
    \langle \Psi, \Phi \rangle =
    \sum_{\substack{i,j=1 \\ i<j}}^V \sum_{n_{i,j}=0}^{N_{i,j}}  
    C_{i,j} \Psi_{i,j}^{(a_{i,j})}(x_{n_{i,j}}^{(a_{i,j})})  
    \Phi_{i,j}^{(a_{i,j})*}(x_{n_{i,j}}^{(a_{i,j})}).
\end{eqnarray}
where elements of $E$ component wavefunction vector $\Phi$ on the discrete graph are defined as
\begin{equation*}
    \left(\Phi_{i,j}^{(a_{i,j})}(x_{n_{i,j}}^{(a_{i,j})})  \right)_{n_{i,j}=0}^{N_{i,j}},
\end{equation*}
for $(i,j)$ such that $C_{i,j}=1$, where $i<j$.

\begin{figure}[t!]
    \includegraphics[width=75mm]{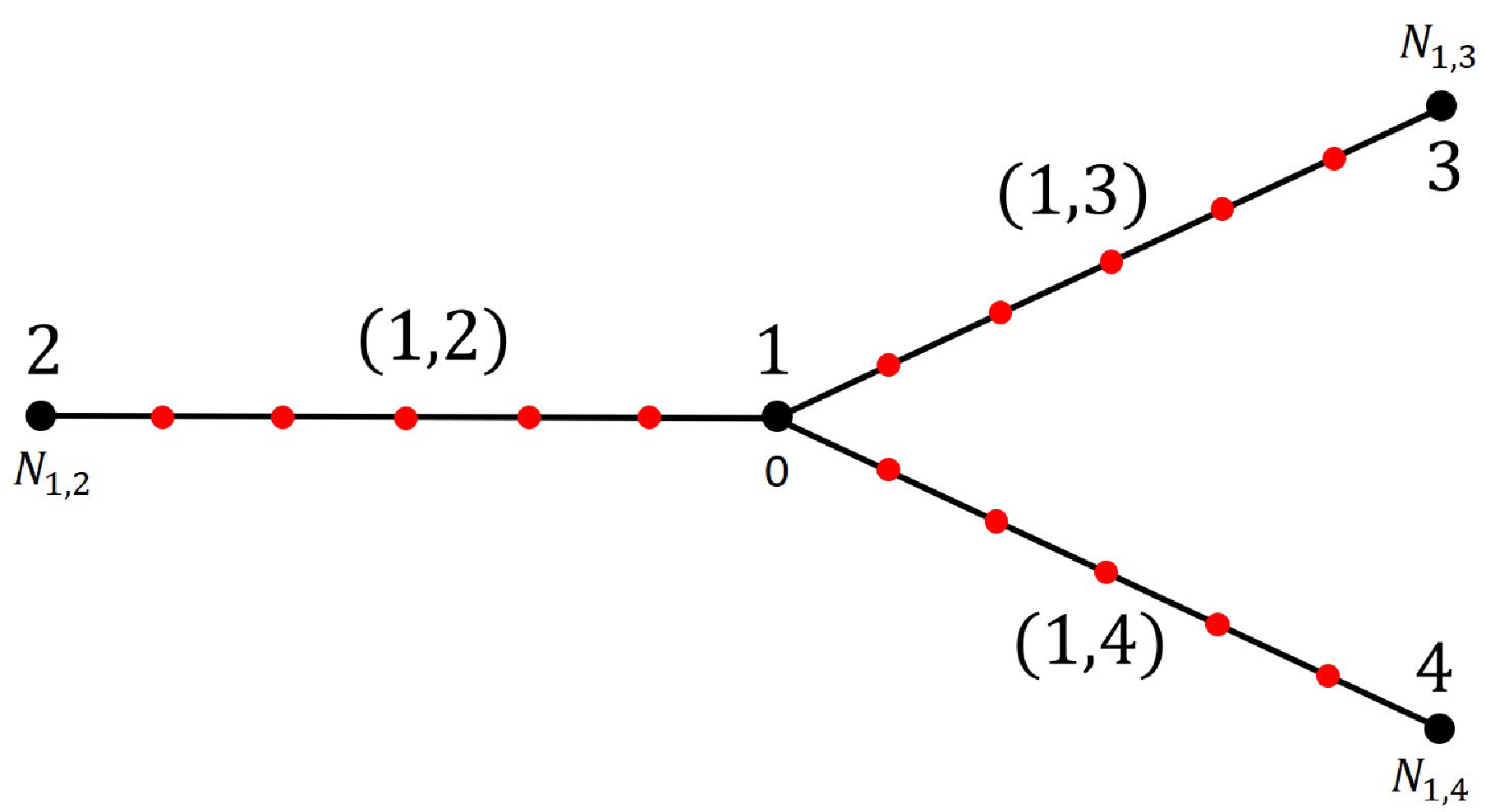}
    \caption{Discrete star graph.}
    \label{fig:star}
\end{figure}

As usual, the discrete Schr\"odinger operator $H_d$  on the graph
corresponds to the negative second discrete derivative.
It is defined component wise on each edge of the
graph via
\begin{align*}
\Psi_{i,j}^{(a_{i,j})}(x_{n_{i,j}}^{(a_{i,j})}) & \mapsto \\ 
-\frac{1}{a_{i,j}^2} &\bigg[\Psi_{i,j}^{(a_{i,j})}(x_{n_{i,j}}^{(a_{i,j})}-a_{i,j}) \\
& - 2\Psi_{i,j}^{(a_{i,j})}(x_{n_{i,j}}^{(a_{i,j})}) + \Psi_{i,j}^{(a_{i,j})}(x_{n_{i,j}}^{(a_{i,j})}+a_{i,j}) \bigg].
\end{align*}

It satisfies
\begin{align}
    \langle H_d\Psi, \Phi\rangle - \langle& \Psi,  H_d\Phi\rangle 
    = \sum_{\substack{i,j=1 \\ i<j}}^V C_{i,j} \bigg[  
    \Psi_{i,j}^{(a_{i,j})}(0) \Phi_{i,j}^{(a_{i,j})*}(a_{i,j}) \nonumber\\  
    & -\Psi_{i,j}^{(a_{i,j})}(a_{i,j}) \Phi_{i,j}^{(a_{i,j})*}(0) \nonumber\\
    & + \Psi_{i,j}^{(a_{i,j})}(L_{i,j}) \Phi_{i,j}^{(a_{i,j})*}(L_{i,j}-a_{i,j}) \nonumber\\ \label{Gambas}
    & - \Psi_{i,j}^{(a_{i,j})}(L_{i,j}-a_{i,j}) \Phi_{i,j}^{(a_{i,j})*}(L_{i,j})
    \bigg].
\end{align}

In fact, $H_d$ is a matrix, and, hence, it is self-adjoint 
if and only if \eqref{Gambas} is zero. Similar as in the previous section, we impose a continuity condition and 
some current conservation rule. The  continuity condition
reads as
\begin{equation}
    \Psi_{i,j}^{(a_{i,j})}(0)=\phi_i, \quad \Psi_{i,j}^{(a_{i,j})}(L_{i,j})=\phi_j,\label{vbc_1}
\end{equation}
for some complex numbers $\phi_{i}$ for all $i<j$ and
$C_{i,j} = 1$.
The discrete version of the current conservation
\begin{align}
\begin{split}
    -&\sum_{j<i}C_{i,j}\left[\Psi_{j,i}^{(a_{j,i})}(L_{j,i})-\Psi_{j,i}^{(a_{j,i})}(L_{j,i}-a_{j,i})\right] \\
    &+\sum_{j>i}C_{i,j}\left[\Psi_{i,j}^{(a_{i,j})}(a_{i,j})-\Psi_{i,j}^{(a_{i,j})}(0)\right] 
     = \lambda_i \phi_i, \label{vbc_2}
\end{split}
\end{align}
for some real parameters $\lambda_i$ for $i$ in $\{1,...,V\}$.

Inspired from Eq.~\eqref{sol1} we use a somehow different linear combination of the fundamental solution in \eqref{sol1} which is adapted to the above continuity and current conservation law similar in \eqref{sol_continuous} as
\begin{align}\label{sol_discrete_graph}
    \Psi_{i,j}^{(a_{i,j})}(x_{n_{i,j}}^{(a_{i,j})}) = & \nonumber\\
    C_{i,j} B_{i,j}  \bigg[ &
    g_+(a_{i,j})^\frac{x_{n_{i,j}}^{(a_{i,j})}}{a_{i,j}}  - 
    g_-(a_{i,j})^\frac{x_{n_{i,j}}^{(a_{i,j})}}{a_{i,j}} \bigg] \\
    +C_{i,j} A_{i,j}  \bigg[ &
    g_+(a_{i,j})^\frac{L_{i,j}-x_{n_{i,j}}^{(a_{i,j})}}{a_{i,j}}  - 
    g_-(a_{i,j})^\frac{L_{i,j}-x_{n_{i,j}}^{(a_{i,j})}}{a_{i,j}} \bigg], \nonumber
\end{align}
for $i<j$, where 
\begin{equation*}
g_{\pm}(a_{i,j})=1+\frac{-k^2 a^2_{i,j} \pm ka_{i,j}\sqrt{k^2 a^2_{i,j}-4}}{2}.    
\end{equation*}

These functions are the exact solution of the discrete Schr\"odinger equation \eqref{eq_graph}. Continuous limit of the solution can be obtained by considering the limit at $a_{i,j}\to 0$ for fixed point $x_{i,j}^{(a_{i,j})}=\alpha_{i,j}$. Similarly, as it is shown in the section~\ref{dse_on_aline}, one obtains
\begin{align}\label{limit_graph}
    \lim_{a_{i,j}\to 0} \Psi_{i,j}^{(a_{i,j})}(\alpha_{i,j}) = & \nonumber\\
    C_{i,j} B_{i,j} & \left[ e^{ik\alpha_{i,j}}-e^{-ik\alpha_{i,j}} \right] \\
    +C_{i,j} A_{i,j} & \left[ e^{ik\left(L_{i,j}-\alpha_{i,j}\right)}-e^{-ik\left(L_{i,j}-\alpha_{i,j}\right)} \right] \nonumber.
\end{align}
The right-hand side of Eq.~\eqref{limit_graph} is analogue of the solution of continuous Schr\"odinger equation on a continuous graph in Eq.~\eqref{sol_continuous}.

\begin{table}\caption{The first five nonzero eigenvalues, $k$ compared with the continuous case on the star graph.}
\centering
\begin{tabular}{l*{4}{|c}r}
                 & Continuous & $a=0.1$     & $a=0.01$   & $a=0.005$ \\
\hline
1            & 1.1799688     & 1.2293914    & 1.1847666  & 1.1823638 \\
2            & 1.6768750     & 1.7771792    & 1.6865131  & 1.6816835 \\
3            & 2.0943951     & 2.0905692    & 2.0943568  & 2.0943855 \\
4            & 2.7486684     & 2.8462967    & 2.7654332  & 2.7570666 \\
5            & 2.8559933     & 2.9088003    & 2.8558962  & 2.8559690
\end{tabular}
\label{tab:eigenvalues_star}
\end{table}

The vertex boundary conditions in Eqs.~\eqref{vbc_1} and \eqref{vbc_2} lead to the following homogeneous system of linear equations:
\begin{align}
A_{i,j} f_{i,j}(N_{i,j}) = \phi_i,\quad
B_{i,j} f_{i,j}(N_{i,j}) = \phi_j, & \\
   \sum_{j<i} C_{i,j}\bigg\{
        A_{i,j} f_{i,j}(1) + B_{i,j} \bigg[  f_{i,j}(N_{i,j}-1)   - & f_{i,j}(N_{i,j})  \bigg]
    \bigg\}  \nonumber\\
    +\sum_{j>i} C_{i,j}\bigg\{
        A_{i,j} \left[  f_{i,j}(N_{i,j}-1)-f_{i,j}(N_{i,j})  \right] + & B_{i,j} f_{i,j}(1)
    \bigg\} \nonumber\\
    =&\lambda_i \phi_i. \label{secular}
\end{align}
where $f_{i,j}(n_{i,j}) = g_+(a_{i,j})^{n_{i,j}}-g_-(a_{i,j})^{n_{i,j}}$.
This is a system of homogeneous equations for $A_{i,j}, \; B_{i,j}$ and $\phi_i$, which has a non-trivial solution when
\begin{eqnarray}\label{seceq}
    \det(M(k))=0,
\end{eqnarray}
where $M(k)$ matrix has the same size and structure as in the section~\ref{cont_graph}.

\subsection{Star branched lattice}

In this subsection, we consider quantum star graph with three discrete edges (see, Fig.~\ref{fig:star}).
For such a graph, $N_{1,j}a_{1,j}=L_{1,j}$ is the length of the edge the $(1,j)$ for $j=2,3,4$.

We choose $\lambda_1=\lambda_2=\lambda_3=\lambda_4=0.$
Then the  continuity conditions \eqref{vbc_1} and the current 
conservation \eqref{vbc_2} are for  $j=2,3,4$
\begin{align}\label{Kebab}
\begin{split}
    &\Psi_{1,j}^{(a_{1,j})}(0)=\phi_1, \quad \Psi_{1,j}^{(a_{1,j})}(L_{1,j})=\phi_j, \\ 
    &\Psi_{1,j}^{(a_{1,j})}(L_{1,j})-\Psi_{1,j}^{(a_{1,j})}(L_{1,j}-a_{1,j})=0, \\ 
    &\sum_{j=2}^4[\Psi_{1,j}^{(a_{1,j})}(a_{1,j})-\Psi_{1,j}^{(a_{1,j})}(0)]=0.
\end{split}
\end{align}

Substitution of the solution in Eq.~\eqref{sol_discrete_graph} into \eqref{Kebab} yields the secular equations
\begin{align}
    &A_{1,j} f_{1,j}(N_{1,j}) = \phi_1,\quad
    B_{1,j} f_{1,j}(N_{1,j}) = \phi_j,\nonumber\\
    &A_{1,j} f_{1,j}(1)+ B_{1,j} \left[ f_{1,j}(N_{1,j}-1)-f_{1,j}(N_{1,j}) \right] = 0,\label{secular_star}\\
    &\sum_{j=2}^4 \left\{
        A_{1,j} \left[f_{1,j}(N_{1,j}-1)-f_{1,j}(N_{1,j})\right] + B_{1,j} f_{1,j}(1)\right\} = 0. \nonumber
\end{align}

The above system of equations has a non-trivial solution, when
\begin{eqnarray}
    \det{(M(k))}=0,
\end{eqnarray}
where the $M(k)$ matrix is the same as the continuous version of the star graph 
\begin{eqnarray}
    M(k)=
    \begin{pmatrix}
        \Lambda^{(1)} & W(k) &0\\
        \Lambda^{(2)} & 0&W(k)\\
        0 & U^{(1)}(k) & U^{(2)}(k)
    \end{pmatrix}.
\end{eqnarray}
The elements of the above matrices are 
\begin{align*}
    &\Lambda^{(1)} = \begin{pmatrix}
        -1 & 0 & 0 & 0 \\
        -1 & 0 & 0 & 0 \\
        -1 & 0 & 0 & 0
    \end{pmatrix},\quad
    \Lambda^{(2)} = \begin{pmatrix}
         0 & -1 &  0 & 0 \\
         0 &  0 & -1 & 0 \\
         0 &  0 &  0 & -1
    \end{pmatrix},\\[1ex]
    &W(k)= \begin{pmatrix}
        f_{1,2}(N_{1,2}) & 0 & 0 \\
        0 & f_{1,3}(N_{1,3}) & 0 \\
        0 & 0 & f_{1,4}(N_{1,4})
    \end{pmatrix},\\[1ex]
    &U^{(1)}(k) = \begin{pmatrix}
        r_{1,2}(N_{1,2}) & r_{1,3}(N_{1,3}) & r_{1,4}(N_{1,4}) \\
        f_{1,2}(1) & 0 & 0 \\
        0 & f_{1,3}(1) & 0 \\
        0 & 0 & f_{1,4}(1)
    \end{pmatrix}, \\[1ex]
    &U^{(2)}(k) = \begin{pmatrix}
        f_{1,2}(1) & f_{1,3}(1) & f_{1,4}(1) \\
        r_{1,2}(N_{1,2}) & 0 & 0 \\
        0 & r_{1,3}(N_{1,3}) & 0 \\
        0 & 0 & r_{1,4}(N_{1,4})
    \end{pmatrix},
\end{align*}
where $r_{1,j}(N_{1,j}) = f_{1,j}(N_{1,j}-1)-f_{1,j}(N_{1,j})$.
This secular equation can be solved numerically to provide eigenvalues of discrete quantum star graph. 
Here one should note that if the lengths of the edges are equal the eigenvalue spectrum becomes degenerate, i.e. has multiple eigenvalues. To avoid such situation, when we solve the secular equation, we choose lengths of the edges as rationally independent.
The first five nonzero eigenvalues are calculated for the considered star graph (see Table~\ref{tab:eigenvalues_star}), where the step size for each edge of the graph is chosen to be equal, $a_{1,j}=a$ and lengths of edges of the graph are chosen as $L_{1,2}=0.8, \; L_{1,3}=1.1, \; L_{1,4}=1.5$. In each column (2-6 columns) in Table \ref{tab:eigenvalues_star} we present nonzero eigenvalues by choosing the different values of step size $a$ for the star graph.
The results show the convergence of solutions to the solutions of the continuous case 
calculated using Eq.~\eqref{eigenvalues_continuous} for the continuous star graph for the same lengths of edges.

\begin{figure}[t!]
    \centering
    \subfloat[\centering Linear (unbranched) conducting polymer lattice.]{{\includegraphics[width=70mm]{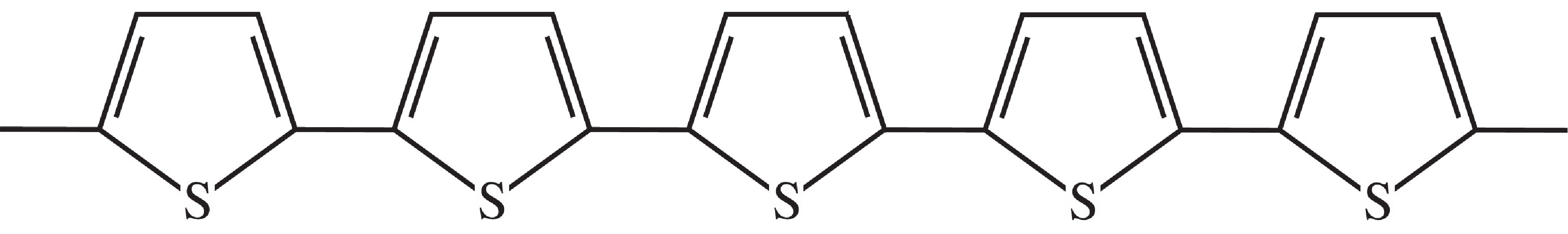} }}
    \quad
    \subfloat[\centering Branched conducting polymer.]{{\includegraphics[width=70mm]{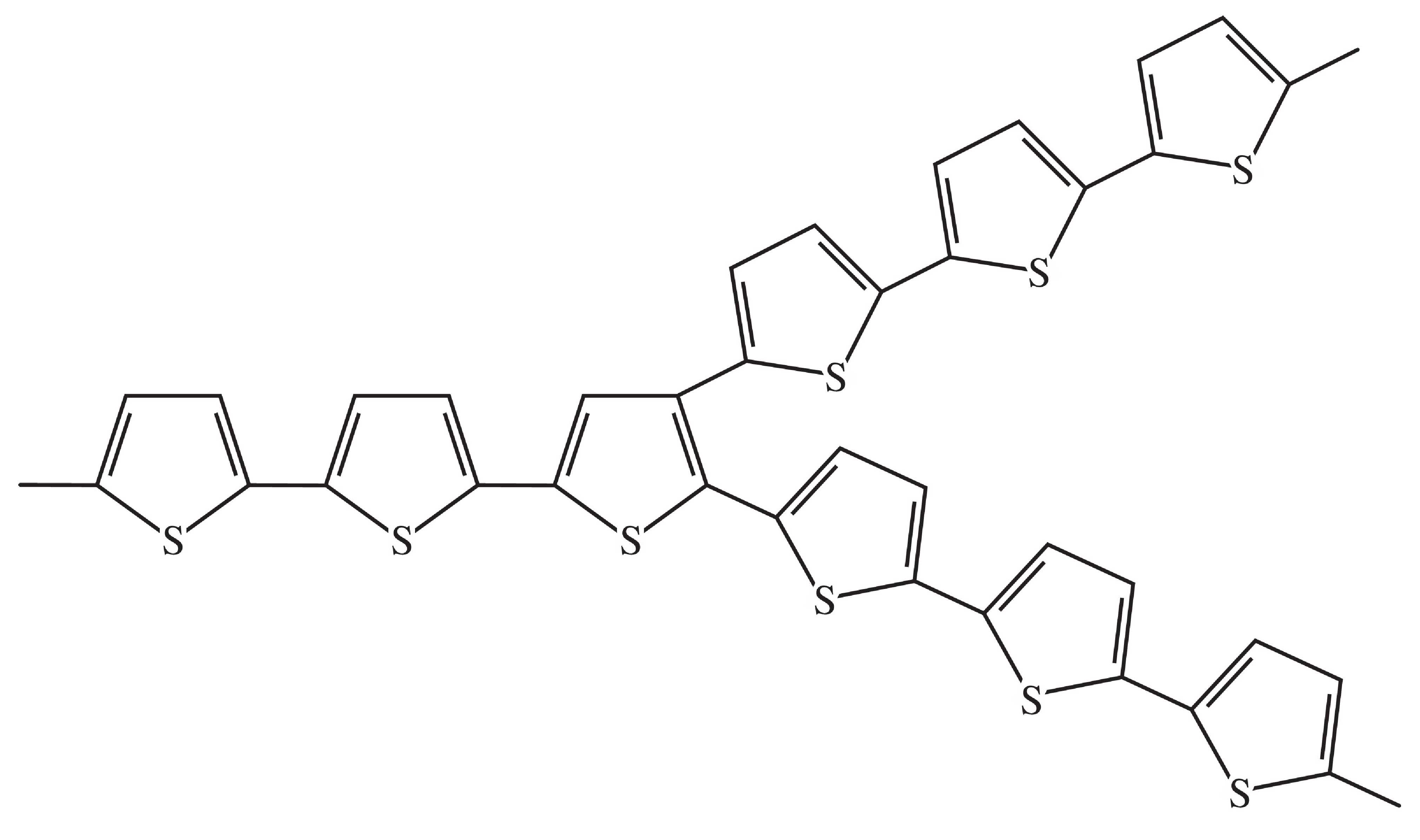} }}
    \caption{Conducting polymer chains.}
    \label{fig:polymer_star}
\end{figure}

\subsection{Possible experimental realization of the model}

Here we briefly discuss the possible practical application of our model to so-called conducting polymers. These latter are the pol-conjugated (polymer) molecular chains exhibiting semiconducting electronic properties and therefore are called organic semiconductors. Conducting polymers are the basic functional materials for organic electronics and so-called polymer based film organic photovoltaics (see e.g., Refs.~\cite{Exciton,DSGE,heeg1,heeg2,heeg3,ak}, for review of conducting polymers and their applications). The structure of conducting polymers presents a discrete lattice (chain) with the hexagonal basic cell (see, Fig.~\ref{fig:polymer_star}a). Charge carrier dynamics in such structure can be described in terms of the discrete Schr\"odinger equation given by Eq.~\eqref{eq2}. An important problem arising in the context of charge dynamics in conducting polymers is modeling electron or exciton (in some cases polarons and solitons) transport with account of its discrete structure. The solution of such a problem allows us to compute the electronic band structure of the material and tuning of its electronic properties. The latter is important for functional optimization of conducting polymers and improving of device performance in organic photovoltaics. Recently, the structures consisting of branched conducting polymer chains attracted much attention in the context of polymer based photovoltaics and organic electronics (see, e.g. Refs \cite{Polaron,Exciton,ak} for review). In the simplest case, such polymer can have a star-branched structure (see Fig.~\ref{fig:polymer_star}b). 
Our model for branched quantum lattice based on the use of discrete Schr\"odinger equation on the metric graph can be effectively used for the description of charge carrier dynamics and computation of the electronic band spectrum in branched conducting polymers \cite{Polaron,Exciton}. Also, charge current, and current-voltage characteristics in the AC driven case can be computed.
Another possible application can be using DSE for modeling of so-called coherent Ising machines, which can be realized using the network states \cite{Huang}. Also the model we proposed can be used for description of chip-based photonic graph states \cite{Lu} and holonomic quantum gates \cite{Li}.

\section{Conclusion}
In this paper, we considered the problem of discrete Schr\"odinger equation on a metric graph, by focusing on its analytical solution and continuum limit. An exact solution of DSE on a finite interval is obtained. Consistency of the result with their continuum counterpart is shown by explicit calculation of the continuum limit of the solution. 
The result is extended to the case of quantum graph of arbitrary topology. Application of the obtained result to the special case of quantum star graph is demonstrated by imposing  the vertex matching conditions given in the form of continuity and Kirchhoff rule. It is shown that the secular equation derived for star graph reproduces its well-know continuum counterpart.
The practical application of the proposed model to branched conducting polymers is discussed.

\section{Acknowledgements}
This work is supported by European Union’s Horizon 2020 research and innovation programme under the Marie Sklodowska-Curie grant agreement ID: 873071, project SOMPATY.
The work of JY and DM is partially supported by the grant of the Innovation Development Agency of the Republic of Uzbekistan (Ref. No. F-2021-440).


\begin{thebibliography}{99}

\bibitem{Pauling} L. Pauling, J. Chem. Phys. {\bf 4}, 673 (1936).
\bibitem{Rud} K. Ruedenberg, C.W. Scherr, J. Chem. Phys. \textbf{21}, 1565 (1953).
\bibitem{Alex} S. Alexander, Phys. Rev. B \textbf{27}, 1541 (1985).
\bibitem{Exner1} P. Exner, P. Seba, P. Stovicek, J. Phys. A: Math. Gen. {\bf 21}, 4009 (1988).
\bibitem{Kost} V. Kostrykin, R. Schrader  J. Phys. A: Math. Gen. {\bf 32}, 595 (1999)
\bibitem{Uzy1} T. Kottos, U. Smilansky, Ann. Phys., {\bf 76}, 274 (1999).
\bibitem{Exner15} P. Exner, H. Kovarik, {\it Quantum waveguides}, (Springer, 2015).
\bibitem{Hul} O. Hul \textit{et al}, Phys. Rev. E {\bf 69}, 056205 (2004).
\bibitem{PTSQGR}  D.U. Matrasulov, J.R. Yusupov, K.K. Sabirov, J. Phys. A {\bf 52}, 155302 (2019).

\bibitem{Grisha} G. Berkolaiko, P. Kuchment, {\it Introduction to Quantum Graphs, Mathematical Surveys and Monographs} AMS (2013).
\bibitem{Kurasov} P. Kurasov, \textit{Spectral Geometry of Graphs}, Springer-Verlag, Berlin (2024).

\bibitem{Jambul} J.R. Yusupov, K.K. Sabirov, M. Ehrhardt, D.U. Matrasulov, Phys. Lett. A {\bf 383}, 2382 (2019).

\bibitem{JM} J. Matrasulov, K. Sabirov, Physica A \textbf{608}, 128279 (2022).
\bibitem{KarimBdG} K.K. Sabirov, J. Yusupov, D. Jumanazarov, D. Matrasulov, Phys. Lett. A {\bf 382}, 2856 (2018).

\bibitem{Alex1} F. Dreisow, M. Heinrich, R. Keil, \textit{et al}, Phys. Rev. Lett., {\bf 105}, 143902 (2010).

\bibitem{PQG2} P. Exner, O. Turek,  J. Phys. A: Math. Theor. {\bf 50} 455201 (2017).
\bibitem{PQG6} E Korotyaev, N Saburova, J. Math. Anal. Appl. {\bf 436} 104 (2016).
\bibitem{Rabinovich} V.Barrera-Figueroa, V.S.Rabinovich, J. Phys. A: Math. Theor. {\bf 50} 215207 (2017).

\bibitem{Grisha2} L. Alon, R. Band, G. Berkolaiko, Commun. Math. Phys., {\bf 362} 909--948 (2018).

\bibitem{PQG8} P. Exner, J. Lipovsk\'y, Phys. Lett. A, {\bf 384} (18), 126390, (2020)


\bibitem{Rezapour1} S. Rezapour, C.T. Deressa, A. Hussain, S. Etemad, R. George, B. Ahmad, Mathematics. \textbf{10}(4), 568 2022.
\bibitem{Rezapour2} D. Baleanu, S. Etemad, H. Mohammadi, S. Rezapour, Communications in Nonlinear Science and Numerical Simulation, \textbf{100}, 105844 (2021).
\bibitem{Rezapour3} S. Etemad, S. Rezapour, Advances in Difference Equations volume \textbf{2020}, 276 (2020).
\bibitem{Rezapour4} D. Baleanu, S.M. Aydogn, H. Mohammadi, S. Rezapour, Alexandria Engineering Journal, \textbf{59}, 3029 (2020).
\bibitem{Rezapour5} H. Khan, K. Alam, H. Gulzar, S. Etemad, S. Rezapour, Mathematics and Computers in Simulation, \textbf{198}, 455 (2022).
\bibitem{Rezapour8} S. Hussain, E.N. Madi, H. Khan, H. Gulzar, S. Etemad, S. Rezapour, M.K.A. Kaabar, Journal of Function Spaces \textbf{9}(23), 4320865 (2022).
\bibitem{Rezapour9} M. Ahmad, A. Zada, M. Ghaderi, R. George, S. Rezapour, Fractal and Fractional \textbf{6}(4), 203 (2022).
\bibitem{Rezapour10} H. Khan, J. Alzabut, A. Shah, Z.-Y. He, S. Etemad, S. Rezapour, A. Zada, Fractals \textbf{31} (04), 2340055 (2023).
\bibitem{Rezapour11} S.M. Aydogan, D. Baleanu, H. Mohammadi, S. Rezapour, Advances in Difference Equations \textbf{2020}, 382 (2020).


\bibitem{DSE2}  A. A  Kvitsinsky, J. Phys. A. {\bf 27},  215 (1994).
\bibitem{DSE1} V. E.Tarasov, Phys. Lett. A, {\bf 380}   68 (2016).
\bibitem{PQG7} E Korotyaev, N Saburova,  Math. Annal. {\bf 377} 723 (2020).
\bibitem{PQG5} E Korotyaev, N Saburova, J. Math. Anal. Appl. {\bf 420} 576 (2014).



\bibitem{Exciton} J.R. Yusupov, Kh.Sh. Matyokubov, K.K. Sabirov, D.U. Matrasulov, Chem. Phys., \textbf{537}, 110861 (2020).
\bibitem{DSGE} M.E. Akramov, J.R. Yusupov, I.N. Askerzade, D.U. Matrasulov, Physica Scripta, \textbf{98}, 115238 (2023).
\bibitem{heeg1} A.J. Heeger, R. Pethig, Philos. Trans. R. Soc. Lond. A \textbf{314}, 17 (1985).
\bibitem{heeg2} A.J. Heeger, Rev. Mod. Phys. \textbf{73}, 681 (2001).
\bibitem{heeg3} A.J. Heeger, S. Kivelson, J.R. Schrieffer, W.-P. Su, Rev. Mod. Phys. \textbf{60}, 781 (1988).
\bibitem{ak} T. Soganci, O. Gumusay, H.C. Soyleyici, M. Ak, Polymer \textbf{134}, 187 (2018).

\bibitem{Polaron} K.K. Sabirov, J.R. Yusupov, Kh.Sh. Matyokubov,  NANOSYSTEMS: Phys., Chem., Math. {\bf 11}, 183 (2020).


\bibitem{Huang} J. Huang, X. Chen, X. Li, J. Wang, AAPPS Bulletin \textbf{33}, 14 (2023).
\bibitem{Lu} B. Lu, L. Liu, J.-Y. Song, K. Wen, Ch. Wang, AAPPS Bulletin \textbf{33}, 7 (2023).
\bibitem{Li} Y. Li, T. Xin, Ch. Qiu, K. Li, G. Liu, J. Li, Y. Wan, D. Lu, Fundamental Research \textbf{3}, 229 (2023).




\end{thebibliography}
\end{document}